\documentclass[journal]{IEEEtran}

\usepackage{ascmac}
\usepackage{amsmath,amssymb}
\usepackage{amsthm}
\usepackage{algorithm,algorithmic}
\usepackage{bm}
\usepackage{mathtools}

\makeatletter
\let\MYcaption\@makecaption
\makeatother

\usepackage{subcaption}

\makeatletter
\let\@makecaption\MYcaption
\makeatother

\theoremstyle{remark}
\newtheorem{remark}{Remark}

\DeclareMathOperator*{\minimize}{minimize~}

\DeclareMathOperator*{\argmin}{arg~min~}

\DeclareMathOperator{\prox}{prox}

\DeclareMathOperator{\supp}{supp}

\newcommand{\abs}[1]{\left\lvert#1\right\rvert}
\newcommand{\norm}[1]{\left\lVert#1\right\rVert}
\newcommand{\paren}[1]{\left(#1\right)}
\newcommand{\sqb}[1]{\left[#1\right]}
\newcommand{\cub}[1]{\left\{#1\right\}}

\usepackage{tikz}
\usetikzlibrary{shapes,arrows,calc}

\definecolor{myblue}{rgb}{0.184,0.392,0.620}
\definecolor{myorange}{rgb}{0.941,0.550,0.078}
\definecolor{mygreen}{rgb}{0,0.560,0}
\definecolor{myred}{rgb}{0.721,0.212,0.118}

\usetikzlibrary{shapes.geometric, arrows, positioning}
\tikzstyle{process} = [rectangle, minimum width=3cm, minimum height=1cm, text centered, text=white]
\tikzstyle{process_v} = [rectangle, minimum width=0.8cm, minimum height=3.0cm, text centered, text=white]
\tikzstyle{arrow} = [very thick,->,>=latex,draw=black!70]

\usepackage{pgfplots}
\pgfplotsset{compat=newest}

\usepackage{cite}

\allowdisplaybreaks[1]

\begin{document}

\title{Loss Function Design for Deep Unfolded Sparse Signal Recovery: Supervised and Unsupervised Learning}

\author{Koshi Nagahisa, Ryo Hayakawa,~\IEEEmembership{Member,~IEEE}, and Youji Iiguni
\thanks{This work was supported by JSPS KAKENHI Grant Number JP24K17277 and the Nakajima Foundation.}%
\thanks{Koshi Nagahisa and Youji Iiguni are with Graduate School of Engineering Science, The University of Osaka, 1-3 Machikaneyama, Toyonaka, Osaka, 560-8531, Japan.}%
\thanks{Ryo Hayakawa is with Institute of Engineering, Tokyo University of Agriculture and Technology, 2-24-16 Naka-cho, Koganei, Tokyo, 184-8588, Japan.}%
}

\markboth{}%
{Loss Function Design for Deep Unfolded Sparse Signal Recovery: Supervised and Unsupervised Learning}


\maketitle

\begin{abstract}
    This paper investigates the impact of loss function design in deep unfolding techniques for sparse signal recovery algorithms. 
    We focus on deep unfolded versions of the fundamental iterative shrinkage thresholding algorithm (ISTA) and the iterative hard thresholding algorithm (IHT). 
    To obtain a guideline for the loss function design, we examine the effect of supervised learning using mean squared error and unsupervised learning using the objective function of the original optimization problem. 
    Our simulation results reveal that the effect of loss function design significantly depends on the convexity of the optimization problem. 
    For convex $\ell_1$-regularized problems, supervised-ISTA achieves better final recovery accuracy but fails to minimize the original objective function, whereas we empirically observe that unsupervised-ISTA converges to a nearly identical solution as conventional ISTA but with accelerated convergence. 
    Conversely, for nonconvex $\ell_0$-regularized problems, both supervised-IHT and unsupervised-IHT converge to better local minima than the original IHT, showing similar performance under the training conditions regardless of the loss function employed.
    However, when the test conditions differ from the training conditions, our results suggest that unsupervised learning offers better robustness to distribution mismatch.
    These findings provide valuable insights into the design of effective deep unfolded networks for sparse signal recovery applications.
\end{abstract}


\section{Introduction} \label{sec:introduction}
\IEEEPARstart{O}{ne} of the fundamental problems in signal processing is the estimation of an unknown vector from low-dimensional linear measurements. 
Such linear inverse problems, known as underdetermined systems, inherently possess an infinite number of solutions without additional constraints. 

Compressed sensing~\cite{CS,Candes2005-ut,Candes2006-eb} is a framework for solving underdetermined linear inverse problems. 
In compressed sensing, we assume that the solution is sparse, i.e., it has few nonzero elements, and estimate the solution by exploiting this sparsity property. 
Such sparse signal recovery problems appear in various fields, including medical imaging~\cite{Lustig2007-vj}, wireless communications~\cite{Hayashi2013-nq,Choi2017-dk}, and control theory~\cite{Nagahara2020-sz}. 

A typical optimization formulation for sparse signal recovery is the $\ell_{1}$ reconstruction problem. 
The optimization problem consists of a differentiable data fidelity term and a non-differentiable $\ell_{1}$ regularization term. 
The proximal gradient method~\cite{PGM1,PGM2} is an efficient algorithm for solving optimization problems with non-differentiable objective functions. 
When applied to the $\ell_{1}$ reconstruction problem, the proximal gradient method yields the iterative shrinkage thresholding algorithm (ISTA)~\cite{ISTA, Combettes2005-mf, Figueiredo2007-bj}. 
Another approach uses the $\ell_{0}$ norm as a regularization term. 
Applying the proximal gradient method to this $\ell_{0}$-regularized problem leads to the iterative hard thresholding algorithm (IHT)~\cite{IHT}. 

The convergence speed and estimation accuracy of iterative algorithms are affected by their parameters, such as step size. 
In order to learn these parameters and improve recovery performance, deep unfolding~\cite{LISTA,Monga2021-ph,Shlezinger2023-zd,Chen2022-ga} has been proposed. 
This approach unfolds the signal flow of the iterative algorithm and learns the parameters within the algorithm by using deep learning techniques. 
The first deep unfolded network is learned ISTA (LISTA)~\cite{LISTA,Chen2018-bq} proposed in the context of sparse coding, where the goal is to find a sparse representation of a given signal. 
In deep unfolding, designing the learnable parameters of the unfolded network is crucial to obtain good recovery performance. 
For example, the learnable parameters in LISTA are the step size and the elements of some matrices in the update equations of ISTA. 
In another method called step LISTA (SLISTA)~\cite{Ablin2019-hu}, the step size of ISTA is learned via deep unfolding to efficiently select dictionary elements and obtain sparse representations in sparse coding problems. 
In the context of sparse signal recovery, trainable ISTA (TISTA)~\cite{TISTA} has been proposed as a deep-unfolded signal recovery method. 
During the training of TISTA, the step sizes of the algorithm are the only learnable parameters. 

In deep unfolding, various functions are used as loss functions for training. 
For example, LISTA uses the mean squared error (MSE) between the output of the network and the estimate obtained by the coordinate descent method (CoD)~\cite{CoD}. 
Since the ideal output for sparse coding does not exist in advance, the estimate obtained by CoD is used as the target vector for the network. 
In SLISTA, the loss function is obtained by substituting the output of the network into the original objective function for sparse coding.
This approach can be regarded as unsupervised learning for the deep unfolded network. 
On the other hand, since TISTA is designed for sparse signal recovery, supervised learning with true signal data is performed during training. 
Specifically, the loss function for TISTA is the MSE between the true signals and their estimates. 

Although both supervised and unsupervised learning can be used in sparse signal recovery when the true signal data are available, the effect of the loss function remains unclear even for simple algorithms such as ISTA and IHT. 
How does the choice of loss function affect the values of learned parameters and the performance of algorithms using them? 
Also, is there a difference in behavior between convex and nonconvex optimization? 

In this paper, we first investigate the effect of loss functions in deep unfolded ISTA. 
In the experiments, we focus on learning step size parameters of ISTA. 
Computer simulations show that ISTA with learned step sizes through supervised learning can achieve better final estimation accuracy than the original ISTA, as it minimizes the error with respect to the true signals during training. 
However, it does not minimize the objective function of the original ISTA in general. 
On the other hand, ISTA with learned step sizes through unsupervised learning can obtain a nearly identical estimate as the original ISTA and its convergence speed is faster than the original ISTA. 
This occurs because the objective function of the original ISTA is used as the loss function in unsupervised learning. 

Moreover, we also examine the impact of loss functions in deep unfolded IHT in nonconvex optimization contexts, using a similar approach to the case of ISTA.
As a result, the learned parameters for IHT enhance the convergence speed compared to the original IHT.
Moreover, in terms of the estimation accuracy, both supervised and unsupervised learning provide more accurate estimates than the original IHT, and the choice of loss function has little effect on the final accuracy under the training conditions.
These phenomena are completely different from the case of ISTA with convex $\ell_{1}$ norm, suggesting that the influence of the loss function depends on the convexity of the original optimization problem for sparse signal recovery.
While the two learning approaches yield similar accuracy under the training conditions, they differ in robustness: unsupervised-IHT generalizes well when the test conditions deviate from the training conditions, whereas supervised-IHT tends to suffer from performance degradation.

The rest of the paper is organized as follows. 
Section~\ref{sec:CS} explains the problem setting of sparse signal recovery and some fundamental algorithms. 
Section~\ref{sec:DU} introduces deep unfolding and several unfolded methods. 
Our motivation and discussion of the loss function are presented in Section~\ref{sec:loss_function}. 
Simulation results of supervised and unsupervised learning are provided in Section~\ref{sec:simulation}. 
Finally, we present our conclusions in Section~\ref{sec:conclusion}. 
\section{Optimization-Based Sparse Signal Recovery} \label{sec:CS}
In this section, we first describe the problem settings of sparse signal recovery. 
We then explain some optimization problems and algorithms for sparse signal recovery. 
\subsection{Sparse Signal Recovery and Sparse Coding}
In compressed sensing, also known as sparse signal recovery, an unknown sparse vector $\bm{x}^{\ast} \in \mathbb{R}^{N}$ is recovered from the linear measurement $\bm{y} \in \mathbb{R}^{M}$ obtained as
\begin{align}
    \bm{y} 
    = 
    \bm{A} \bm{x}^{\ast} + \bm{v}, \label{eq:measurement_model}
\end{align}
where $\bm{A} \in \mathbb{R}^{M \times N}$ ($M < N$) is the known measurement matrix and $\bm{v} \in \mathbb{R}^{M}$ is the measurement noise vector. 

The sparse coding problem can be formulated in the same way as in~\eqref{eq:measurement_model}. 
In sparse coding, however, the main goal is to find a sparse representation of a given signal $\bm{y}$ using a known dictionary $\bm{A}$. 
While the mathematical model appears identical, there is a fundamental difference in the problem setup. 
In sparse signal recovery, the sparse signal $\bm{x}^{\ast}$ exists first and the measurement $\bm{y}$ is obtained from it through the measurement process. 
On the other hand, in sparse coding, we begin with the signal $\bm{y}$ and aim to decompose it into a linear combination of dictionary atoms (columns of $\bm{A}$) with as few nonzero coefficients as possible.
\subsection{$\ell_{0}$ Reconstruction and $\ell_{1}$ Reconstruction}
Since $M < N$, there are an infinite number of solutions even if there is no measurement noise. 
One approach to the problem is to take advantage of the sparsity of the solution. 
A naive approach is the $\ell_{0}$ reconstruction problem given by
\begin{align}
    \hat{\bm{x}} 
    =
    \argmin_{\bm{x} \in \mathbb{R}^{N}}
    \cub{
        \frac{1}{2} \norm{\bm{y} - \bm{A} \bm{x}}^{2}_{2}
        + \lambda \norm{\bm{x}}_{0}
    }, \label{eq:opt_L0}
\end{align}
where $\norm{\cdot}_{2}$ is the $\ell_{2}$ norm of the vector. 
Here, $\norm{\bm{x}}_{0}$ is called $\ell_{0}$ norm of the vector $\bm{x} = \sqb{x_{1}, \dotsc, x_{N}}^{\top}$ and defined as
\begin{align}
    \norm{\bm{x}}_{0} 
    = 
    \abs{\supp (\bm{x})}, 
\end{align}
where $\supp (\bm{x}) = \cub{n \in \cub{1, \dotsc, N} \mid x_{n} \neq 0 }$ is the support set of $\bm{x}$ and $\abs{\cdot}$ here denotes the number of elements in the set. 
From the definition, $\norm{\bm{x}}_{0}$ represents the number of nonzero elements of the vector $\bm{x}$. 
The first term $\frac{1}{2} \norm{\bm{y} - \bm{A} \bm{x}}_{2}^{2}$ in the objective function of~\eqref{eq:opt_L0} is the data fidelity term and represents the difference between $\bm{y}$ and $\bm{A} \bm{x}$. 
$\lambda$ ($>0$) is the regularization parameter to adjust which of the two terms is more important. 
As $\lambda$ increases, the influence of the regularization term in~\eqref{eq:opt_L0} becomes stronger, and the solution is more likely to be sparse. 
However, the optimization problem in~\eqref{eq:opt_L0} is a combinatorial optimization problem due to the discreteness and nonconvexity of $\norm{\bm{x}}_0$. 
Thus, it is computationally difficult to find the exact solution of~\eqref{eq:opt_L0}, especially for a large-scale problem.

To tackle the difficulty of the $\ell_{0}$ reconstruction, the $\ell_1$ reconstruction problem given by 
\begin{align}
    \hat{\bm{x}} 
    = 
    \argmin_{\bm{x} \in \mathbb{R}^{N}}
    \cub{
        \frac{1}{2} \norm{\bm{y} - \bm{A} \bm{x}}_{2}^{2}
        + \lambda \norm{\bm{x}}_{1}
    } \label{eq:opt_L1}
\end{align}
is often considered as a relaxed convex optimization problem. 
Here, the $\ell_{0}$ norm in~\eqref{eq:opt_L0} is replaced with the convex $\ell_1$ norm, which is defined as 
\begin{align}
    \norm{\bm{x}}_{1} 
    = 
    \sum_{n = 1}^{N} \abs{x_{n}}. 
\end{align}
The objective function of the $\ell_{1}$ reconstruction is continuous and convex, and hence the local optima of the problem are also the global optima.
\subsection{Proximal Gradient Method}
Even with the convex relaxation by the $\ell_1$ norm, the objective function of the optimization problem in~\eqref{eq:opt_L1} is not differentiable.
Consequently, the standard gradient descent method cannot be applied directly. 
The proximal gradient method~\cite{PGM1,PGM2} is one of the optimization algorithms to solve such optimization problems.
This method is an algorithm for the unconstrained minimization problem 
\begin{align}
    \minimize_{\bm{x} \in \mathbb{R}^{N}} 
    \cub{ f(\bm{x}) + g(\bm{x}) }, \label{eq:opt_PGM}
\end{align}
where $f: \mathbb{R}^{N} \to \mathbb{R}$ is a differentiable convex function and $g: \mathbb{R}^{N} \to \mathbb{R}$ is a convex function (not necessarily differentiable). 
The update equations of the proximal gradient method for the problem in~\eqref{eq:opt_PGM} are given by
\begin{subequations}
    \begin{align}
        \bm{r}^{(t)} 
        &=
        \bm{x}^{(t)} - \alpha \nabla f(\bm{x}^{(t)}), \label{eq:PGM1} \\
        \bm{x}^{(t+1)} 
        &=
        \prox_{\alpha g}(\bm{r}^{(t)}), \label{eq:PGM2}
    \end{align}
\end{subequations}
where $\bm{x}^{(0)} \in \mathbb{R}^{N}$ is the initial value and $\alpha$ ($>0$) denotes the step size. 
The update in~\eqref{eq:PGM1} is the gradient descent step using the gradient $\nabla f(\bm{x})$. 
The update in~\eqref{eq:PGM2} is based on the proximity operator of the function $g: \mathbb{R}^{N} \to \mathbb{R}$, which is defined as
\begin{align}
    \prox_{\alpha g}(\bm{x})
    =
    \argmin_{\bm{u} \in \mathbb{R}^{N}}
    \cub{
        \alpha g(\bm{u}) 
        + \frac{1}{2} \norm{\bm{u} - \bm{x}}_{2}^{2}
    }. \label{eq:prox}
\end{align}

The algorithm of the proximal gradient method is summarized in Algorithm~\ref{alg:PGM}. 
\begin{algorithm}[t]
    \begin{algorithmic}[1]
        \caption{Proximal gradient method}
        \label{alg:PGM}
        \REQUIRE $\bm{x}^{(0)}$, $\alpha$
        \ENSURE $\bm{x}^{(t)}$
        \WHILE{the stop condition is not satisfied}
            \STATE $\bm{r}^{(t)} = \bm{x}^{(t)} - \alpha \nabla f(\bm{x}^{(t)})$
            \STATE $\bm{x}^{(t+1)} = \prox_{\alpha g}(\bm{r}^{(t)})$
            \STATE $t \leftarrow t + 1$
        \ENDWHILE
    \end{algorithmic}
\end{algorithm}
In the proximal gradient method, an appropriate step size ensures convergence to the optimal solution. 
Suppose that the gradient $\nabla f$ of the function $f$ is Lipschitz continuous on $\mathbb{R}^{N}$ and the Lipschitz constant is $L$. 
That is, $L$ is the smallest value that satisfies
\begin{align}
    \norm{\nabla f(\bm{x}_{1}) - \nabla f(\bm{x}_{2})}_{2}
    \le 
    L \norm{\bm{x}_{1} - \bm{x}_{2}}_{2} \quad (\forall \bm{x}_{1}, \bm{x}_{2} \in \mathbb{R}^{N}).
\end{align}
By setting the step size $\alpha$ to satisfy the condition 
\begin{align}
    0 < \alpha \le \frac{1}{L}, \label{eq:convergence_condition}
\end{align}
the sequence $\{\bm{x}^{(t)}\}$ obtained by Algorithm~\ref{alg:PGM} converges to the optimal solution $\hat{\bm{x}}$~\cite{Beck2009-uk, Parikh2014-sz}. 
\subsection{ISTA}
The proximal gradient method for the $\ell_{1}$ reconstruction problem in~\eqref{eq:opt_L1} is called ISTA~\cite{ISTA}. 
The optimization problem in~\eqref{eq:opt_L1} is obtained by setting $f(\bm{x}) = \frac{1}{2} \norm{\bm{A} \bm{x} - \bm{y}}_{2}^{2}$ and $g(\bm{x}) = \lambda \norm{\bm{x}}_{1}$ in~\eqref{eq:opt_PGM}, and we can apply the proximal gradient method directly to the $\ell_{1}$ reconstruction problem. 
In this case, the gradient of function $f(\bm{x})$ is given by $\nabla f(\bm{x}) = \bm{A}^{\top} (\bm{A} \bm{x} - \bm{y})$. 
Moreover, from~\eqref{eq:prox}, the proximity operator of $g(\bm{x}) = \lambda \norm{\bm{x}}_{1}$ in~\eqref{eq:PGM2} can be written as
\begin{align}
    \sqb{\prox_{\alpha g} (\bm{x})}_{n}
    &=
    S_{\lambda \alpha}(x_{n}) \\
    &\coloneqq
    \begin{cases}
       x_{n} - \lambda \alpha & (x_{n} > \lambda \alpha) \\
       0 & (- \lambda \alpha \le x_{n} \le \lambda \alpha) \\
       x_{n} + \lambda \alpha & (x_{n} < - \lambda \alpha),
   \end{cases} \label{eq:soft_thresholding}
\end{align}
where $\sqb{\cdot}_{n}$ denotes the $n$-th element of the vector. 
$S_{\lambda \alpha} (\cdot)$ is called the soft-thresholding function. 
If the input of $S_{\lambda \alpha}(\cdot)$ is a vector, we apply the function in~\eqref{eq:soft_thresholding} to each element of the vector. 
In summary, from~\eqref{eq:PGM1} and~\eqref{eq:PGM2}, the update equations of ISTA can be written as 
\begin{subequations}
    \begin{align}
        \bm{r}^{(t)} 
        &=
        \bm{x}^{(t)} - \alpha \bm{A}^{\top} (\bm{A} \bm{x}^{(t)} - \bm{y}), \label{eq:ISTA1} \\
        \bm{x}^{(t+1)} 
        &=
        S_{\lambda \alpha} (\bm{r}^{(t)}). \label{eq:ISTA2}
    \end{align}
\end{subequations}

The algorithm of ISTA is summarized in Algorithm~\ref{alg:ISTA}. 
\begin{algorithm}[t]
    \begin{algorithmic}[1]
        \caption{ISTA}
        \label{alg:ISTA}
        \REQUIRE $\bm{x}^{(0)}$, $\alpha$, $\bm{y}$, $\bm{A}$
        \ENSURE $\bm{x}^{(t)}$
        \WHILE{the stop condition is not satisfied}
            \STATE $\bm{r}^{(t)} = \bm{x}^{(t)} - \alpha \bm{A}^{\top} (\bm{A} \bm{x}^{(t)} - \bm{y})$
            \STATE $\bm{x}^{(t+1)} = S_{\lambda \alpha} (\bm{r}^{(t)})$
            \STATE $t \leftarrow t + 1$
        \ENDWHILE
    \end{algorithmic}
\end{algorithm}
From~\eqref{eq:convergence_condition}, if the step size $\alpha$ satisfies 
\begin{align}
    0
    <
    \alpha
    \le
    \frac{1}{\lambda_{\text{max}} (\bm{A}^{\top} \bm{A})}, 
\end{align}
ISTA converges to the optimal solution, where $\lambda_{\text{max}} (\bm{A}^{\top} \bm{A})$ denotes the largest eigenvalue of $\bm{A}^{\top} \bm{A}$. 
\subsection{IHT}
Next, we consider the proximal gradient method for the $\ell_{0}$ reconstruction in~\eqref{eq:opt_L0}, though the optimization problem is nonconvex. 
When $g(\bm{x}) = \lambda \norm{\bm{x}}_{0}$, the proximity operator of the function $g$  can be written as\footnote{Strictly speaking, $\prox_{\alpha g}(x_{n})$ is set-valued at the boundary $\abs{x_{n}} = \sqrt{2 \lambda \alpha}$, i.e., $\prox_{\alpha g}(x_{n}) = \cub{0, x_{n}}$ for $\abs{x_{n}} = \sqrt{2 \lambda \alpha}$. 
In this case, we keep $x_{n}$ in~\eqref{eq:hard_thresholding}.
This convention has negligible impact on the performance of the algorithm in practice because this event occurs only when the input $x_{n}$ is exactly equal to the threshold. }
\begin{align}
    \sqb{\prox_{\alpha g}(\bm{x})}_{n} 
    &=
    H_{\lambda \alpha} (x_{n}) \\
    &\coloneqq
    \begin{cases}
        x_{n} & (\abs{x_{n}} \ge \sqrt{2 \lambda \alpha})\\
        0 & (\abs{x_{n}} < \sqrt{2 \lambda \alpha})
    \end{cases}. \label{eq:hard_thresholding}
\end{align}
The function $H_{\lambda \alpha} (\cdot)$ is called the hard thresholding function. 
The update equations obtained by formally applying the proximal gradient method to the optimization problem in~\eqref{eq:opt_L0} are given by 
\begin{subequations}
    \begin{align}
        \bm{r}^{(t)} 
        &=
        \bm{x}^{(t)} - \alpha \bm{A}^{\top} (\bm{A} \bm{x}^{(t)} - \bm{y}), \label{eq:IHT1} \\
        \bm{x}^{(t+1)} 
        &=
        H_{\lambda \alpha} (\bm{r}^{(t)}). \label{eq:IHT2}
    \end{align}
\end{subequations}
This iterative algorithm is called IHT~\cite{IHT}\footnote{There are two common formulations of IHT: the sparsity-constrained version using $\tilde{H}_k(\bm{z})$, which retains the $k$ largest-magnitude elements and requires the sparsity level $k$ to be specified in advance, and the regularization-based version using $H_{\lambda\alpha}(\cdot)$ derived as the proximal operator of $\lambda\|\bm{x}\|_0$.
In our simulations, we adopt the latter so that ISTA and IHT can be treated within a unified proximal gradient framework, enabling a systematic comparison across convex ($\ell_1$) and nonconvex ($\ell_0$) settings.} and is shown in Algorithm~\ref{alg:IHT}.
\begin{algorithm}[t]
    \begin{algorithmic}[1]
        \caption{IHT}
        \label{alg:IHT}
        \REQUIRE $\bm{x}^{(0)}$, $\alpha$, $\bm{y}$, $\bm{A}$
        \ENSURE $\bm{x}^{(t)}$
        \WHILE{the stop condition is not satisfied}
            \STATE $\bm{r}^{(t)} = \bm{x}^{(t)} - \alpha \bm{A}^{\top} (\bm{A} \bm{x}^{(t)} - \bm{y})$
            \STATE $\bm{x}^{(t+1)} = H_{\lambda \alpha} (\bm{r}^{(t)})$
            \STATE $t \leftarrow t + 1$
        \ENDWHILE
    \end{algorithmic}
\end{algorithm}
It should be noted that the optimization problem of the $\ell_{0}$ reconstruction is nonconvex and there can be multiple local minima. 
Thus, it does not necessarily converge to a global minimum in general. 
\section{Parameter Learning via Deep Unfolding} \label{sec:DU}
Although ISTA described in Section~\ref{sec:CS} is guaranteed to converge to the optimal solution with appropriate step sizes, the convergence speed heavily depends on the value of the step size. 
Thus, determining appropriate step sizes for each iteration is desirable to achieve faster convergence. 
One promising approach for the convergence acceleration is deep unfolding~\cite{LISTA,Monga2021-ph,Shlezinger2023-zd,Chen2022-ga}, which utilizes machine learning techniques for neural networks. 
This section describes deep unfolding and its application to sparse signal recovery and sparse coding. 
\subsection{Deep Unfolding}
Deep unfolding is a technique for learning the parameters of iterative algorithms. 
In deep unfolding, we first unfold the signal flow graph of a conventional iterative algorithm in the time direction. 
Then, we regard the unfolded graph as a feedforward neural network. 
Finally, the parameters of the algorithm are trained by using deep learning techniques such as backpropagation and stochastic gradient descent. 

An example of signal flow graph for an iterative algorithm is shown in Fig.~\ref{fig:deep_unfolding}(\subref{subfig:signal_flow}). 
\begin{figure*}[t]
    \centering
    \begin{minipage}[c]{0.3\linewidth}
        \centering
        \scalebox{0.6}{
            \begin{tikzpicture}[node distance=1.6cm]
                \node (procA) [process, fill=myblue!60, font=\Large] {A};
                \node (procB) [process, below of=procA, fill=myblue!80, font=\Large] {B};
                \node (procC) [process, below of=procB, fill=myblue!100, font=\Large] {C};
                
                \draw [arrow] ++(0,1.2) node[above, font=\Large] {Input} -- (procA.north);
                \draw [arrow] (procA) -- (procB);
                \draw [arrow] (procB) -- (procC);
                \draw [arrow] (procC.south) -- ++(0,-0.5) -- ++(2.5,0) -- ++(0,4.2) -- (procA.east);
                \draw [arrow] (procC.south) -- ++(0,-1.2) node[below, font=\Large] {Output};

            \end{tikzpicture}
        }
        \subcaption{Signal flow of iterative algorithms.}
        \label{subfig:signal_flow}
    \end{minipage}
    \begin{minipage}[c]{0.69\linewidth}
        \centering
        \scalebox{0.8}{
            \begin{tikzpicture}[node distance=1.4cm]
                \node (procA1) [process_v, fill=myblue!60, font=\Large] {A};
                \node (procB1) [process_v, right of=procA1, fill=myblue!80, font=\Large] {B};
                \node (procC1) [process_v, right of=procB1, fill=myblue!100, font=\Large] {C};
                
                \node (circleA1) [circle, fill=myorange!80, inner sep=2mm] at (procA1.south) [yshift=4mm] {};
                \node (circleB1) [circle, fill=myorange!80, inner sep=2mm] at (procB1.south) [yshift=4mm] {};
                \node (circleC1) [circle, fill=myorange!80, inner sep=2mm] at (procC1.south) [yshift=4mm] {};
                
                % ...
                \node (proc_space) [process_v, right of=procC1, fill=white, text=black!70, font=\large] {$\cdots$};
                
                \node (procA2) [process_v, right of=proc_space, fill=myblue!60, font=\Large] {A};
                \node (procB2) [process_v, right of=procA2, fill=myblue!80, font=\Large] {B};
                \node (procC2) [process_v, right of=procB2, fill=myblue!100, font=\Large] {C};
                
                \node (circleA2) [circle, fill=myorange!80, inner sep=2mm] at (procA2.south) [yshift=4mm] {};
                \node (circleB2) [circle, fill=myorange!80, inner sep=2mm] at (procB2.south) [yshift=4mm] {};
                \node (circleC2) [circle, fill=myorange!80, inner sep=2mm] at (procC2.south) [yshift=4mm] {};

                \node (trainParam) [below right=20mm and -3mm of circleC2, font=\large, text=myorange!80] {trainable parameters};
                \draw [arrow, draw=myorange!80] (trainParam) -- (circleC2);
    
                \node (loss) [right=8mm of procC2, font=\large] {Loss function};
    
                \draw [arrow] (procA1) -- (procB1) node[midway, above, yshift=0.5mm] {};
                \draw [arrow] (procB1) -- (procC1) node[midway, above, yshift=0.5mm] {};
                \draw [arrow] (procC1) -- (proc_space) node[midway, above, yshift=0.5mm] {};
                
                \draw [arrow] (proc_space) -- (procA2) node[midway, above, yshift=0.5mm] {};
                \draw [arrow] (procA2) -- (procB2) node[midway, above, yshift=0.5mm] {};
                \draw [arrow] (procB2) -- (procC2) node[midway, above, yshift=0.5mm] {};
                \draw [arrow] (procC2) -- (loss) node[midway, above, yshift=0.5mm] {};
                
                \draw [arrow, dashed] (loss) -- ++(0,-2.5) -- ++(-10.9,0) node[midway, below, font=\large] {update by backpropagation} -- (circleA1.south);
                \draw [arrow, dashed] (8.4,-2.4)-- (circleC2.south);
                \draw [arrow, dashed] (7.0,-2.4) -- (circleB2.south);
                \draw [arrow, dashed] (5.6,-2.4) -- (circleA2.south);
                \draw [arrow, dashed] (2.8,-2.4) -- (circleC1.south);
                \draw [arrow, dashed] (1.4,-2.4) -- (circleB1.south);
            \end{tikzpicture}
        }
        \subcaption{Unfolded signal flow.}
        \label{subfig:unfolded_signal_flow}
    \end{minipage}
    \caption{An example of signal flow for an iterative algorithm and its unfolded version.}
    \label{fig:deep_unfolding}
\end{figure*}
Subprocesses A, B, and C in the figure represent the update equations at each iteration of the algorithm.
Fig.~\ref{fig:deep_unfolding}(\subref{subfig:unfolded_signal_flow}) shows the unfolded signal flow graph obtained by expanding the iterative process in the time direction.
From the figure, we can see that the structure of the unfolded signal flow graph is similar to feedforward neural networks.
Thus, if the subprocesses are differentiable with respect to their parameters, we can apply deep learning techniques such as backpropagation and stochastic gradient descent to train the parameters of the iterative algorithm. 
\subsection{LISTA}
Deep unfolding was first introduced in~\cite{LISTA}, where an ISTA-based network called LISTA was proposed to improve the performance of ISTA for sparse coding. 
The update equations of the original ISTA in Algorithm~\ref{alg:ISTA} can be summarized as 
\begin{align}
    \bm{x}^{(t+1)} 
    = 
    S_{\lambda \alpha} ((\bm{I}_{N} - \alpha \bm{A}^{\top} \bm{A}) \bm{x}^{(t)} + \alpha \bm{A}^{\top} \bm{y}), \label{eq:ISTA_summary}
\end{align}
where $\bm{I}_{N} \in \mathbb{R}^{N \times N}$ is the identity matrix. 
Letting $\bm{W}_{x} = \bm{I}_{N} - \alpha {\bm{A}^{\top} \bm{A}}$ and $\bm{W}_{y} = \alpha \bm{A}^{\top}$, we can rewrite~\eqref{eq:ISTA_summary} as 
\begin{align}
    \bm{x}^{(t+1)} 
    = 
    S_{\lambda \alpha} (\bm{W}_{x} \bm{x}^{(t)} + \bm{W}_{y} \bm{y}). \label{eq:LISTA}
\end{align}
By considering $(\alpha, \bm{W}_{x}, \bm{W}_{y})$ as the trainable parameters of the network, we can regard the update equation in~\eqref{eq:LISTA} as a layer of a neural network.
Note that $\bm{W}_{x}$ and $\bm{W}_{y}$ are uniquely determined by $\alpha$ and $\bm{A}$ as shown in~\eqref{eq:ISTA_summary}.
LISTA nevertheless treats them as free trainable parameters because it was originally designed for the sparse coding setting where the dictionary $\bm{A}$ is not assumed to be known or fixed.

The parameters in each layer of LISTA can be learned by minimizing a loss function via stochastic gradient descent with backpropagation.
The loss function using data $\bm{y}_{1}, \dotsc, \bm{y}_{N_{b}} \in \mathbb{R}^{M}$ is defined as 
\begin{align}
    \mathcal{L}(\theta)
    =
    \frac{1}{N_{b}} \sum_{i = 1}^{N_{b}}
    \norm{\hat{\bm{x}}_{i} - \bm{x}_{\text{CoD},i}}_{2}^{2}, \label{eq:LISTA_loss}
\end{align}
where $\theta=\cub{\bm{W}_{x}, \bm{W}_{y}, \alpha}$ denotes the set of all trainable parameters and $N_{b}$ is the minibatch size. 
In~\eqref{eq:LISTA_loss}, $\hat{\bm{x}}_{i}$ is the output of the network for the training data $\bm{y}_{i}$ and $\bm{x}_{\text{CoD},i}$ is the estimate by CoD~\cite{CoD}. 
Since LISTA is intended for sparse coding problems, the true value $\bm{x}_{i}^{\ast}$ does not exist in principle, and hence the estimate obtained by CoD is used instead as the target vector. 
The parameters are trained so that the network output $\hat{\bm{x}}$ is close to $\bm{x}_{\text{CoD}}$. 

Another important variant is analytic LISTA (ALISTA)~\cite{Liu2019-gh}, which reduces the number of trainable parameters by analytically determining the weight matrix $\bm{W}$ and learning only the scalar step size $\alpha_t$ and threshold $\lambda_t$ at each iteration.
The update equations of ALISTA are given by
\begin{subequations}
    \begin{align}
        \bm{r}^{(t)} &= \bm{x}^{(t)} - \alpha_t \bm{W} (\bm{A} \bm{x}^{(t)} - \bm{y}), \\
        \bm{x}^{(t+1)} &= S_{\lambda_t}(\bm{r}^{(t)}),
    \end{align}
\end{subequations}
where $\bm{W}$ is fixed and obtained by solving an optimization problem independent of the training data.
With supervised learning, ALISTA achieves exponential convergence while being more parameter-efficient than LISTA.
\subsection{SLISTA}
Although LISTA improves the performance of the original ISTA, the computational cost of the training is high for large-scale problems because we need to learn the matrices $\bm{W}_{x}$ and $\bm{W}_{y}$. 
In another approach named SLISTA~\cite{Ablin2019-hu}, we keep the matrix $\bm{A}$ fixed, and only the step sizes of the original ISTA are learned via deep unfolding. 
The update equation for SLISTA is given by
\begin{subequations}
    \begin{align}
        \bm{r}^{(t)} 
        &=
        \bm{x}^{(t)} - \alpha_{t} \bm{A}^{\top} (\bm{A} \bm{x}^{(t)} - \bm{y}), \label{eq:SLISTA1} \\
        \bm{x}^{(t+1)} 
        &=
        S_{\lambda \alpha_{t}} (\bm{r}^{(t)}). \label{eq:SLISTA2}
    \end{align}
\end{subequations}
The parameters of SLISTA are the step sizes at each iteration, i.e., $\theta = \cub{\alpha_{t}}$ ($t=0, 1, \dotsc, T$), where $T$ is the maximum number of iterations. 

The parameters of SLISTA are learned by using the loss function
\begin{align}
    \mathcal{L}(\theta)
    =
    \frac{1}{N_{b}} \sum_{i = 1}^{N_{b}} 
    \cub{
        \frac{1}{2} \norm{\bm{y}_{i} - \bm{A} \hat{\bm{x}}_{i}}_{2}^{2}
        + \lambda \norm{\hat{\bm{x}}_{i}}_{1}
    }. \label{eq:loss_SLISTA}
\end{align}
In~\eqref{eq:loss_SLISTA}, the measurement data $\bm{y}_{i}$ and the corresponding estimate $\hat{\bm{x}}_{i}$ are substituted into the objective function of the $\ell_{1}$ reconstruction in~\eqref{eq:opt_L1} for each minibatch. 
Training with this loss function can be regarded as unsupervised learning because the true value $\bm{x}_{i}^{\ast}$ is not used in the loss function, which is preferable in the context of sparse coding. 
\subsection{TISTA}
TISTA~\cite{TISTA} has been proposed as a trainable iterative algorithm for sparse signal recovery. 
The update equations of TISTA are given by
\begin{subequations}
    \begin{align}
        \bm{r}^{(t)} 
        &=
        \bm{x}^{(t)} + \alpha_{t} \bm{W} (\bm{y} - \bm{A} \bm{x}^{(t)}), \label{eq:TISTA1} \\
        v_{t}^{2} 
        &=
        \max \cub{\frac{\norm{\bm{y} - \bm{A} \bm{x}^{(t)}}_{2}^{2} - M \sigma^{2}}{\text{trace}(\bm{A}^{\top} \bm{A})}, \varepsilon }, \label{eq:TISTA2} \\
        \tau_{t}^{2} 
        &=
        \frac{v_{t}^{2}}{N} (N + (\alpha_{t}^{2} - 2 \alpha_{t}) M) 
        + \frac{\alpha_{t}^{2} \sigma^{2}}{N} \text{trace}(\bm{W} \bm{W}^{\top}), \label{eq:TISTA3} \\
        \bm{x}^{(t+1)} 
        &=
        \eta_{\text{MMSE}} (\bm{r}^{(t)}; \tau_{t}^{2}), \label{eq:TISTA4}
    \end{align}
\end{subequations}
where $\alpha_{t} \in \mathbb{R}$ ($t=0,1,\dotsc $) is the trainable step size. 
The matrix $\bm{W} = \bm{A}^{\top} (\bm{A} \bm{A}^{\top})^{-1}$ in~\eqref{eq:TISTA1} is the Moore-Penrose pseudoinverse of the matrix $\bm{A}$. 
The real constant $\varepsilon$ ($>0$) in \eqref{eq:TISTA2} is a sufficiently small value to prevent the estimate of the variance from being non-positive. 
$\sigma^{2}$ is the variance of the measurement noise $\bm{v}$. 
The $\text{trace}(\cdot)$ denotes the trace of the matrix. 
The function $\eta_{\text{MMSE}}$ in~\eqref{eq:TISTA4} is the minimum mean squared error (MMSE) estimation function that performs the denoising of the Gaussian noise with variance $\tau_{t}^{2}$ on the basis of the conditional expectation~\cite{TISTA}.

The loss function used for TISTA is given by
\begin{align}
    \mathcal{L}(\theta)
    =
    \frac{1}{N_b} \sum_{i=1}^{N_{b}}
    \norm{\hat{\bm{x}}_{i} - \bm{x}_{i}^{\ast}}_{2}^{2}, \label{eq:loss_TISTA}
\end{align}
where $\theta = \cub{\alpha_{t}}$ is a set of the trainable parameters. 
Additionally, by including some parameters of the MMSE function as learnable parameters, we can train them based on the data. 
The loss function is the MSE of the estimate $\hat{\bm{x}}_{i}$. 
Since the loss function includes the training data of the true value $\bm{x}_{i}^{\ast}$, the parameter training based on~\eqref{eq:loss_TISTA} can be regarded as supervised learning. 
Unlike the case of sparse coding, the true value $\bm{x}_{i}^{\ast}$ exists in sparse signal recovery in principle. 

Incremental training is used for TISTA to address the vanishing gradient problem, which may cause minimal updates of the parameters of the network.
In incremental training for deep unfolding, we first set the total number of layers to $T = 1$, and the loss function is evaluated by using the output of the $1$-layer network for each minibatch.
After updating the parameters a predetermined number of times in the inner loop, the value of $T$ is incremented in the outer loop, up to a maximum value $T_{\text{max}}$.
\subsection{IHT-based Methods}
Deep unfolding has also been applied to IHT in an analogous manner to LISTA~\cite{Wang2016-do}.
Just as LISTA replaces the soft thresholding operator in ISTA with a learned counterpart, IHT-based deep unfolding replaces the hard thresholding operator with a learned one.
A learning approach analogous to that of LISTA has been considered for training.
Variants with layer-specific (non-shared) weight matrices have also been proposed to improve the recovery performance~\cite{Xin2016-gj}.
\section{Impacts of Loss Function in Deep Unfolding} \label{sec:loss_function}
\subsection{Motivation and Research Questions}
Deep unfolding techniques have achieved remarkable success in sparse signal recovery. 
However, one fundamental aspect has received limited attention: the role of the loss function in influencing the behavior and the performance of these methods. 
This issue becomes particularly critical when contrasting supervised and unsupervised learning, as the two frameworks reflect fundamentally different perspectives on what constitutes a ``good'' solution. 

Motivated by this gap, we investigate the following research questions:
\begin{enumerate}
    \item How do supervised and unsupervised loss functions influence the convergence properties of deep unfolded algorithms?
    \item Does the impact of the loss function differ between convex problems (ISTA with $\ell_1$ regularization) and nonconvex problems (IHT with $\ell_0$ regularization)?
\end{enumerate}
\subsection{Loss Function Design for Deep Unfolding}
As described in Section~\ref{sec:DU}, various loss functions have been employed in different deep unfolding-based methods. 
One reason is that the aims of learning in the literature differ between sparse signal recovery and sparse coding. 
In sparse signal recovery, there is a ground truth sparse signal $\bm{x}^{\ast}$ that we aim to recover, and the quality of recovery can be evaluated by comparing the estimate with this ground truth. 
In contrast, sparse coding does not have a unique ``true'' sparse representation\textemdash the objective is merely to find a representation that is sparse while accurately reconstructing the original signal $\bm{y}$. 
This distinction becomes particularly important when designing loss functions for learning-based approaches, as supervised learning requires a ground truth that exists in sparse signal recovery but not necessarily in sparse coding. 
Although both supervised and unsupervised learning can be used for sparse signal recovery, there has been limited investigation into how the design of the loss function affects learning outcomes and the learned parameters. 

\subsection{Algorithms Considered in This Study}
In this study, we focus on sparse signal recovery and examine the effect of loss functions for deep unfolded ISTA and IHT, as simple examples of recovery algorithms. 
Specifically, we consider ISTA with varying step sizes as
\begin{subequations}
    \begin{align}
        \bm{r}^{(t)} 
        &=
        \bm{x}^{(t)} - \alpha_{t} \bm{A}^{\top} (\bm{A} \bm{x}^{(t)} - \bm{y}), \\
        \bm{x}^{(t+1)} 
        &= 
        S_{\lambda \alpha_{t}} (\bm{r}^{(t)}),
    \end{align}
\end{subequations}
and IHT with varying step sizes as
\begin{subequations}
\begin{align}
    \bm{r}^{(t)} 
    &=
    \bm{x}^{(t)} - \alpha_{t} \bm{A}^{\top} (\bm{A} \bm{x}^{(t)} - \bm{y}), \\
    \bm{x}^{(t+1)} 
    &= 
    H_{\lambda \alpha_{t}} (\bm{r}^{(t)}), 
\end{align}
\end{subequations}
where $\alpha_{t}$ is the trainable parameter at each iteration. 
As the loss function, we consider the function 
\begin{align}
    \mathcal{L}(\theta)
    =
    \frac{1}{N_{b}N} 
    \sum_{i=1}^{N_{b}} 
    \norm{\hat{\bm{x}}_{i} - \bm{x}_{i}^{\ast}}_{2}^{2} \label{eq:loss_supervised}
\end{align}
for supervised learning and the function 
\begin{align}
    \mathcal{L}(\theta)
    =
    \frac{1}{N_{b}N} 
    \sum_{i=1}^{N_{b}} 
    \cub{
        \frac{1}{2} \norm{\bm{y}_{i} - \bm{A} \hat{\bm{x}}_{i}}_{2}^{2}
        + \lambda R(\hat{\bm{x}}_{i})
    } \label{eq:loss_unsupervised}
\end{align}
for unsupervised learning, where $\theta = \cub{\alpha_{t}}$ is a set of trainable parameters and $N_{b}$ is the minibatch size. 
For ISTA, the regularization term $R(\cdot)$ is set to $R(\bm{x}) = \norm{\bm{x}}_{1}$, while for IHT, it is set to $R(\bm{x}) = \norm{\bm{x}}_{0}$. 
To stabilize the training process, we slightly modify the hard thresholding function $H_{\lambda \alpha_{t}}(\cdot)$ in IHT by introducing a small positive constant $\varepsilon$ as 
\begin{align}
    H_{\lambda \alpha_{t}}(x) 
    = 
    \begin{cases}
        x & (\abs{x} \ge \sqrt{2 \lambda \alpha_{t}+ \varepsilon}) \\
        0 & (\abs{x} < \sqrt{2 \lambda \alpha_{t}+ \varepsilon})
    \end{cases}. \label{eq:modified_hard_thresholding}
\end{align}
For backpropagation through the hard thresholding function, we use the standard automatic differentiation provided by PyTorch~\cite{Paszke2019-vd}.
Although $H_{\lambda\alpha_t}(\cdot)$ is discontinuous at $\abs{x} = \sqrt{2\lambda\alpha_t + \varepsilon}$, it is differentiable almost everywhere.
Since the probability that the input $x$ takes exactly the threshold value is zero in practice, we use this almost-everywhere derivative in the training.

\begin{remark}[A property of unsupervised learning in deep unfolding] \label{rem:unsup_property}
In unsupervised learning, the loss function $\mathcal{L}(\theta)$ in~\eqref{eq:loss_unsupervised} is the empirical average of the original objective function evaluated at the final iteration.
Thus, minimizing the training loss directly corresponds to minimizing the original objective function.
Since any constant sequence $\theta_\alpha = \cub{\alpha, \ldots, \alpha}$ belongs to the space $\Theta = \cub{\cub{\alpha_t}_{t=0}^{T-1} : \alpha_t > 0}$ of trainable step size sequences, a global minimizer $\theta^*$ of $\mathcal{L}(\theta)$ satisfies $\mathcal{L}(\theta^*) \leq \mathcal{L}(\theta)$ for all $\theta \in \Theta$, including $\theta_\alpha$ for any $\alpha > 0$.
In other words, the ideal learned step size sequence is guaranteed to achieve a lower value of (the sum of) the original objective function than any fixed step size.
Note that this argument assumes a global minimizer is obtained, whereas training yields only an approximate solution in practice.
Furthermore, this is a statement about the training loss, and a separate analysis would be needed to extend it to the test loss.
\end{remark}

\section{Simulation Results} \label{sec:simulation}
\subsection{Simulation Setup}
The training data are generated by
\begin{align}
    \bm{y}_{i} 
    = 
    \bm{A} \bm{x}_{i}^{\ast} + \bm{v}_{i} 
    \quad (i = 1, \dotsc, N_{b}) 
\end{align}
for each minibatch. 
In all simulations, we set $N = 300$ and $M = 210$. 
The nonzero elements of the sparse vector $\bm{x}_{i}^{\ast}$ are independent and identically distributed (i.i.d.) standard Gaussian variables, and the probability that an entry is nonzero is set to $p = 0.1$. 
The measurement matrix $\bm{A}$ is composed of i.i.d.\ Gaussian variables with zero mean and variance $1/N$.
The noise vector $\bm{v}_{i}$ is composed of i.i.d.\ Gaussian variables with zero mean and variance $\sigma_{v}^{2}$. 
The variance $\sigma_{v}^{2}$ is set so that the signal-to-noise ratio (SNR), defined as $10 \log_{10} (\sigma_{x}^{2} / \sigma_{v}^{2})$, equals $20$~dB, where $\sigma_{x}^{2} = p$. 
The regularization parameter $\lambda$ is set to $0.05$ for ISTA and $0.01$ for IHT. 
The initial value of the algorithms is $\bm{x}_{0} = \bm{0}$  for both algorithms. 
The small value $\varepsilon$ in~\eqref{eq:modified_hard_thresholding} is fixed to $10^{-10}$.

In the training, we adopt incremental training with a fixed number of iterations $T_{\text{max}} = 120$. 
We set the initial value of the step size $\alpha_{t}$ ($t=0,1,\dotsc,T_{\text{max}}-1$) to $1/L$, where $L$ is the Lipschitz constant of $\nabla f (\bm{x}) = \bm{A}^{\top} \paren{\bm{A} \bm{x} - \bm{y}}$ and given by the largest eigenvalue of $\bm{A}^{\top} \bm{A}$. 
More precisely, we use $L$ obtained by averaging the Lipschitz constants over $100$ different measurement matrices. 
The training parameters are updated by using the Adam optimizer~\cite{Kingma2015-xy}. 
The learning rate of the optimizer is set to $5.0 \times 10^{-3}$ for ISTA and $1.0 \times 10^{-3}$ for IHT. 
The minibatch size for a single parameter update is $N_{b} = 50$ and the number of parameter updates is $100$ for each stage of the incremental training. 
After each parameter update, we enforce nonnegativity by setting any negative step size parameter to zero. 
This ensures all learned step sizes remain valid throughout training. 

In the test, we evaluate the performance for $100$ different measurement matrices $\bm{A}$ and $100$ different original signals $\bm{x}^{\ast}$ for each $\bm{A}$. 
The distributions of $\bm{x}^{\ast}, \bm{A}, \bm{v}$ are the same as those used in the training, because the aim of this study is to investigate the effect of the loss function on the performance of deep unfolding. 
\subsection{Simulation Results}
\subsubsection{ISTA}
\begin{figure}[tp]
    \centering
    \begin{minipage}[c]{1.0\linewidth}
        \centering
        \includegraphics[height=50mm]{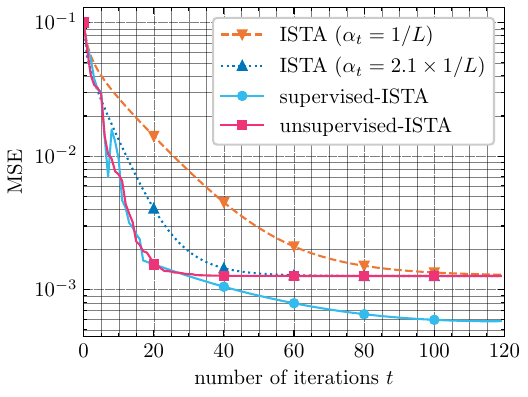}
        \subcaption{MSE performance}
        \label{subfig:MSE_ISTA}
    \end{minipage}
    \begin{minipage}[c]{1.0\linewidth}
        \centering
        \includegraphics[height=50mm]{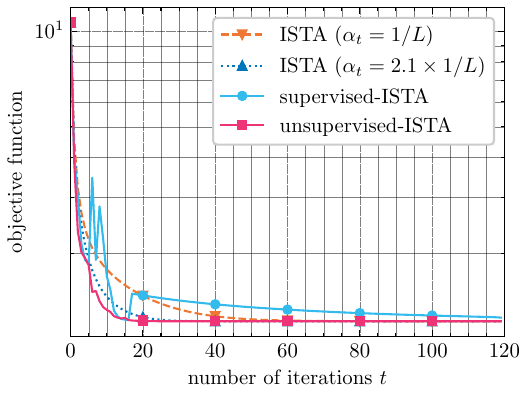}
        \subcaption{Objective function}
        \label{subfig:Objective_ISTA}
    \end{minipage}
    \begin{minipage}[c]{1.0\linewidth}
        \centering
        \includegraphics[height=50mm]{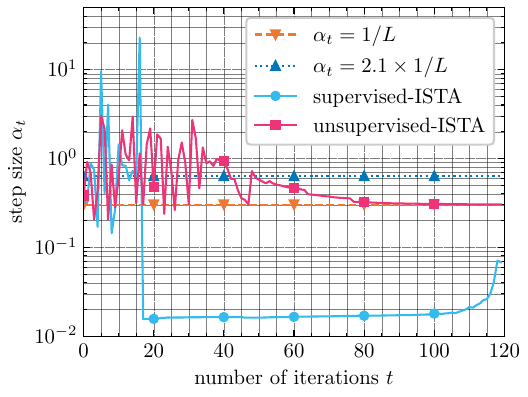}
        \subcaption{Step size}
        \label{subfig:StepSize_ISTA}
    \end{minipage}
    \caption{Simulation results of ISTA ($N = 300$, $M = 210$).}
    \label{fig:ISTA_results}
\end{figure}

We first show the results of ISTA with the learned step size obtained by deep unfolding. 
Figs.~\ref{fig:ISTA_results}(\subref{subfig:MSE_ISTA}) and~\ref{fig:ISTA_results}(\subref{subfig:Objective_ISTA}) show the mean squared error (MSE) performance and objective function value of ISTA with the following step sizes, respectively: 
\begin{itemize}
    \item $\alpha_{t} = 1/L$ (``ISTA ($\alpha_{t} = 1/L$)'')
    \item $\alpha_{t} = 2.1 \times 1/L$ (``ISTA ($\alpha_{t} = 2.1 \times 1/L$)'')
    \item $\alpha_{t}$ obtained by supervised learning (``supervised-ISTA'')
    \item $\alpha_{t}$ obtained by unsupervised learning (``unsupervised-ISTA'')
\end{itemize}
The value $2.1$ was obtained from preliminary experiments as the empirical maximum value at which ISTA converges most rapidly without divergence in most simulations. 
From Fig.~\ref{fig:ISTA_results}(\subref{subfig:MSE_ISTA}), we observe that supervised-ISTA achieves better final MSE than other methods. 
This is because supervised learning utilizes the true value of the unknown vector $\bm{x}_{i}^{\ast}$ and learns to approximate it as closely as possible at the final iteration.
In other words, supervised-ISTA effectively learns a mapping from $\bm{y}$ to $\bm{x}^{\ast}$, correcting the bias inherent in the solution $\hat{\bm{x}}$ caused by the $\ell_1$ regularization. 
As a result, the learned parameters drive the algorithm toward $\bm{x}^{\ast}$ rather than the minimizer of the $\ell_1$ regularized optimization problem.
In fact, from Fig.~\ref{fig:ISTA_results}(\subref{subfig:Objective_ISTA}), we can see that the step size learned by supervised learning fails to minimize the objective function. 
In contrast, for unsupervised-ISTA, both the MSE and objective function values converge to nearly identical values as the original ISTA, but in fewer iterations. 
As discussed in Remark~\ref{rem:unsup_property}, since the unsupervised loss coincides with the original objective function, a globally optimal step size sequence for a fixed number of iterations achieves a lower training loss than any fixed step size. 
Moreover, since incremental training optimizes the step sizes at each stage $T = 1, 2, \ldots, T_{\text{max}}$, this property applies not only at the final iteration but also at intermediate stages, where the learned step sizes are optimized to minimize the objective function after fewer iterations.
Although global optimality and the absence of a train-test gap are not guaranteed in practice, this result provides a theoretical basis consistent with the observed behavior.

We then show the step size $\alpha_{t}$ of each method in Fig.~\ref{fig:ISTA_results}(\subref{subfig:StepSize_ISTA}). 
From Fig.~\ref{fig:ISTA_results}(\subref{subfig:StepSize_ISTA}), we observe that the step size of supervised-ISTA takes large values in the first $20$ iterations, then continues to take small values. 
In contrast, the step size of unsupervised-ISTA exhibits a zigzag pattern around the dotted line given by $\alpha_{t} = 2.1 \times 1 / L$ in the first $40$ iterations. 
This zigzag pattern is also observed for other deep unfolded methods~\cite{TISTA,Takabe2019-yc,Takabe2022-us,Matsuda2025-js}.
\subsubsection{IHT}
\begin{figure}[tp]
    \centering
    \begin{minipage}[c]{1.0\linewidth}
        \centering
        \includegraphics[height=50mm]{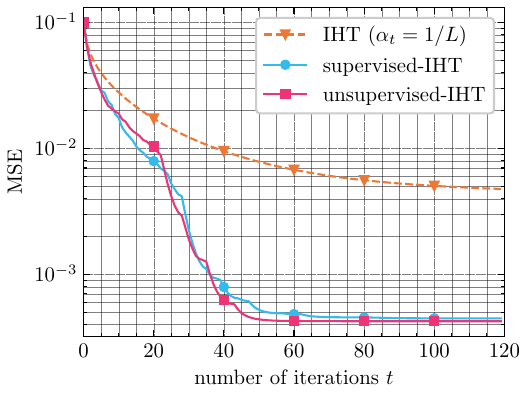}
        \subcaption{MSE performance}
        \label{subfig:MSE_IHT}
    \end{minipage}
    \begin{minipage}[c]{1.0\linewidth}
        \centering
        \includegraphics[height=50mm]{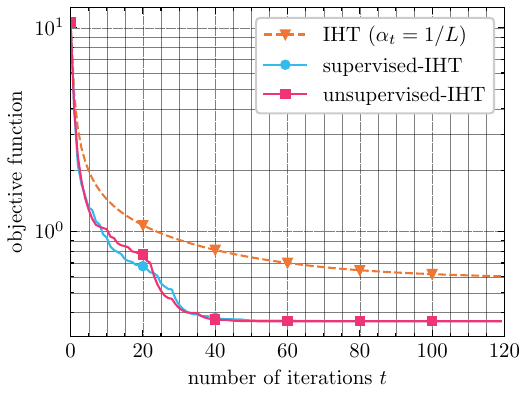}
        \subcaption{Objective function}
        \label{subfig:Objective_IHT}
    \end{minipage}
    \begin{minipage}[c]{1.0\linewidth}
        \centering
        \includegraphics[height=50mm]{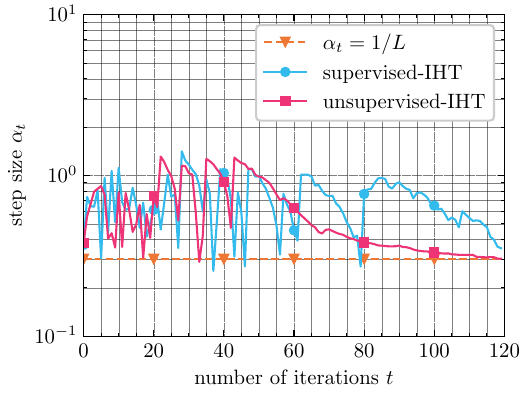}
        \subcaption{Step size}
        \label{subfig:StepSize_IHT}
    \end{minipage}
    \caption{Simulation results of IHT ($N = 300$, $M = 210$).}
    \label{fig:IHT_results}
\end{figure}

We then show the results of deep unfolded-IHT.
Figs.~\ref{fig:IHT_results}(\subref{subfig:MSE_IHT}) and~\ref{fig:IHT_results}(\subref{subfig:Objective_IHT}) show the MSE performance and objective function values of IHT with the following step sizes, respectively: 
\begin{itemize}
    \item $\alpha_{t} = 1/L$ (``IHT ($\alpha_{t} = 1/L$)'')
    \item $\alpha_{t}$ obtained by supervised learning (``supervised-IHT'')
    \item $\alpha_{t}$ obtained by unsupervised learning (``unsupervised-IHT'')
\end{itemize}
From Fig.~\ref{fig:IHT_results}(\subref{subfig:MSE_IHT}), we can see that unsupervised-IHT achieves better MSE than the original IHT, in contrast to unsupervised-ISTA. 
This is partly because the objective function of IHT is nonconvex, and the final MSE is affected by the local minima. 
The parameters obtained by deep unfolding presumably enable IHT to converge to better local minima than the original IHT.
This is also supported by the objective function values in Fig.~\ref{fig:IHT_results}(\subref{subfig:Objective_IHT}), where both supervised-IHT and unsupervised-IHT converge to lower values than the original IHT.
A possible mechanism is suggested by the learned step size schedule shown in Fig.~\ref{fig:IHT_results}(\subref{subfig:StepSize_IHT}), where the learned step sizes tend to take larger values than the fixed step size $1/L$ used in the original IHT.
Such larger step sizes may help the algorithm escape from poor local minima of the nonconvex objective function, potentially leading to convergence to better solutions.
This also explains why supervised-IHT and unsupervised-IHT achieve similar MSE performance.
Since unsupervised learning already provides a large improvement over the original IHT by escaping poor local minima, little room remains for supervised learning to offer additional MSE improvement.
This is in contrast to the ISTA case, where unsupervised-ISTA converges to essentially the same solution as the original ISTA, leaving room for supervised learning to achieve further MSE improvement by correcting the bias inherent in the $\ell_1$-regularized solution.
We also note that the above results are obtained under the zero initialization $\bm{x}_0 = \bm{0}$, which is the standard choice for IHT and is used consistently at both training and test time.
\subsubsection{Robustness to Distribution Mismatch}
 
We evaluate the generalization performance of the learned step sizes when the test conditions differ from the training conditions.
Specifically, we vary the sparsity $p$ and the SNR while keeping all other settings the same as in the training.
The training conditions ($p = 0.1$ and $\mathrm{SNR} = 20$~dB) are indicated by dashed vertical lines in the following figures.

\begin{figure}[tp]
    \centering
    \begin{minipage}[c]{1.0\linewidth}
        \centering
        \includegraphics[height=50mm]{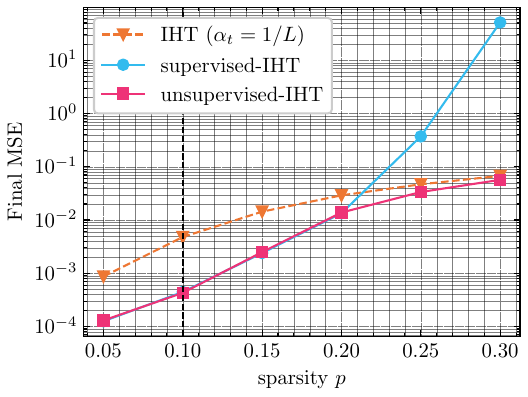}
        \subcaption{Sparsity}
        \label{subfig:generalization_IHT_sparsity}
    \end{minipage}
    \begin{minipage}[c]{1.0\linewidth}
        \centering
        \includegraphics[height=50mm]{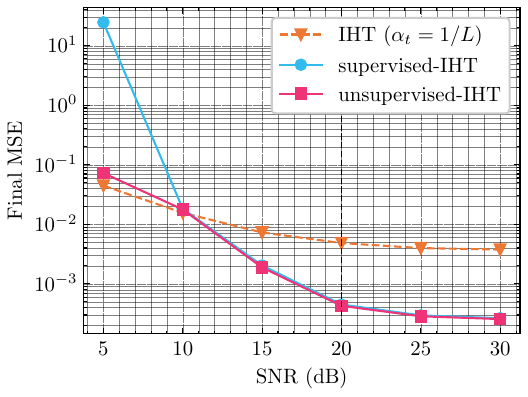}
        \subcaption{SNR}
        \label{subfig:generalization_IHT_SNR}
    \end{minipage}
    \caption{Generalization performance of IHT under distribution mismatch. The dashed vertical lines indicate the training conditions ($p = 0.1$ and $\mathrm{SNR} = 20$~dB).}
    \label{fig:IHT_generalization_results}
\end{figure}

Figs.~\ref{fig:ISTA_generalization_results} and~\ref{fig:IHT_generalization_results} show the final MSE as a function of sparsity $p$ and SNR for ISTA and IHT, respectively.
For ISTA, both supervised-ISTA and unsupervised-ISTA generalize well across a wide range of conditions.
As expected, the MSE increases as the problem becomes harder (i.e., higher sparsity or lower SNR), whereas it decreases in the easier direction; neither method exhibits catastrophic failure.

For IHT, unsupervised-IHT outperforms the original IHT ($\alpha_{t} = 1/L$) in most tested conditions, demonstrating stable generalization.
In contrast, supervised-IHT suffers from significant performance degradation when the conditions deviate substantially from the training point: as shown in Figs.~\ref{fig:IHT_generalization_results}(\subref{subfig:generalization_IHT_sparsity}) and~\ref{fig:IHT_generalization_results}(\subref{subfig:generalization_IHT_SNR}), its MSE increases sharply at high sparsity and low SNR, eventually far exceeding that of the original IHT.
These results suggest that the step sizes learned by supervised-IHT are highly tuned to the training conditions and do not transfer well to different settings.

Overall, unsupervised learning tends to exhibit better robustness to distribution mismatch than supervised learning, particularly for IHT.
This property may be advantageous in practical scenarios where a mismatch between the training and test distributions is unavoidable.
A detailed analysis of the mechanism behind the degradation of supervised-IHT is left as future work.

\begin{figure}[tp]
    \centering
    \begin{minipage}[c]{1.0\linewidth}
        \centering
        \includegraphics[height=50mm]{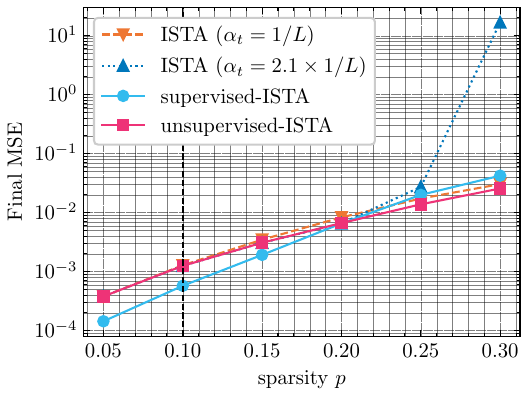}
        \subcaption{Sparsity}
        \label{subfig:generalization_ISTA_sparsity}
    \end{minipage}
    \begin{minipage}[c]{1.0\linewidth}
        \centering
        \includegraphics[height=50mm]{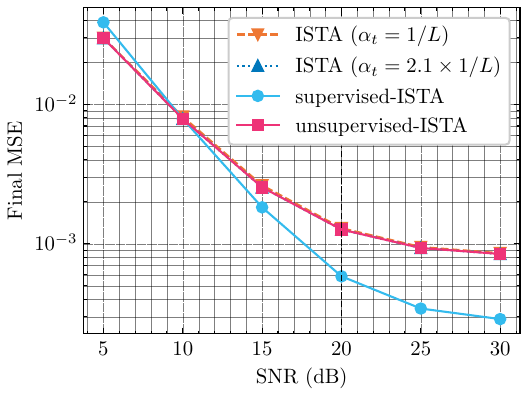}
        \subcaption{SNR}
        \label{subfig:generalization_ISTA_SNR}
    \end{minipage}
    \caption{Generalization performance of ISTA under distribution mismatch. The dashed vertical lines indicate the training conditions ($p = 0.1$ and $\mathrm{SNR} = 20$~dB).}
    \label{fig:ISTA_generalization_results}
\end{figure}
\subsection{Summary}
We summarize the results and discuss the difference between supervised and unsupervised learning in deep unfolded sparse signal recovery. 
In supervised learning using the loss function in~\eqref{eq:loss_supervised}, the goal of the training is to obtain an estimate close to the ground truth $\bm{x}_{i}^{\ast}$. 
Simulation results show that the final MSE of supervised-ISTA is lower than that of the original ISTA. 
However, since the solution $\hat{\bm{x}}_{i}$ obtained by ISTA is usually different from the ground truth $\bm{x}_{i}^{\ast}$, supervised-ISTA does not necessarily converge to the same estimate as ISTA. 
On the other hand, in unsupervised learning with~\eqref{eq:loss_unsupervised}, we use the objective function of the original optimization problem as the loss function during training. 
Thus, unsupervised-ISTA converges to a nearly identical estimate $\hat{\bm{x}}_{i}$ faster than the original ISTA. 
As for IHT, both supervised and unsupervised learning achieve similar levels of accuracy, and both methods converge to more accurate local minima compared to the original IHT. 
This is because the objective function of IHT is nonconvex, and the final MSE is affected by the local minima. 
Furthermore, in terms of robustness to distribution mismatch, unsupervised-IHT generalizes well across a wide range of test conditions, whereas supervised-IHT degrades sharply when the conditions deviate substantially from those used during training.
In summary, the effect of the loss function in deep unfolding depends on the convexity of the original optimization problem, and unsupervised learning tends to exhibit better generalization robustness, particularly for nonconvex problems.

From these findings, we provide the following practical design guidelines for deep unfolded networks based on optimization problems:
\begin{itemize}
    \item \textbf{Convex problems} (e.g., $\ell_1$-regularized sparse recovery): The choice of loss function has a significant impact.
    If the goal is to minimize the original optimization objective, unsupervised learning is preferable, as it converges to the same solution as the original algorithm with accelerated convergence.
    If the goal is to recover the true sparse signal with smaller MSE, supervised learning is more effective, as it directly learns a proper mapping to the true signal. 
    \item \textbf{Nonconvex problems} (e.g., $\ell_0$-regularized sparse recovery): Both supervised and unsupervised learning achieve similar MSE performance, both significantly better than the original algorithm.
    Unsupervised learning is particularly notable, as it improves MSE by helping the algorithm escape poor local minima, without requiring ground-truth training data.
    In addition, unsupervised learning exhibits better robustness to distribution mismatch, making it a more reliable choice when the test conditions may differ from the training conditions.
\end{itemize}
\section{Conclusion} \label{sec:conclusion}
In this study, we have focused on the difference between supervised and unsupervised learning for deep unfolded sparse signal recovery. 
Specifically, we have applied deep unfolding to the fundamental iterative algorithms of ISTA and IHT. 
In the training of the parameters, we have considered two loss functions corresponding to supervised and unsupervised learning in order to evaluate the effect of the loss function on the performance of the algorithms. 
Our simulation results suggest that the property of the trained parameters in deep unfolding significantly depends on the loss function and the convexity of the original optimization problem. 
These results provide useful insights into how the design of the loss function can affect the behavior of deep unfolded algorithms, which may guide future developments in sparse signal recovery and inverse problems. 

Future work includes a similar investigation for other proximal splitting algorithms~\cite{Combettes2011-id} and the optimization problem with nonconvex $\ell_{p}$ regularization ($0<p<1$)~\cite{Wen2018-lu}.
Extending the supervised vs.\ unsupervised comparison framework to other deep unfolding architectures would also be an important direction.
An extension to more complicated problems such as image restoration would also be an interesting topic.
%
%
\bibliographystyle{IEEEtran}
\bibliography{myBib}

\end{document}